\documentclass[twocolumn]{aa}
\usepackage{graphicx}
\usepackage[authoryear]{natbib}
\usepackage{txfonts}
\begin{document}
   \title{Surface imaging of late-type contact
   binaries II: 
   H$\alpha$ 6563$\AA$ emission in AE Phoenicis  and YY Eridani 
   \thanks{based on observations collected at the European Southern 
			 Observatory, La Silla, Chile}
			 }
              

   \author{O. Vilhu
          \inst{1}
          \and
          C. Maceroni\inst{2}
          }

   \offprints{O. Vilhu}

   \institute{Observatory, Box 14, FIN-00014 University of Helsinki, Finland\\
              \email{osmi.vilhu@helsinki.fi}
         \and
             INAF- Osservatorio Astronomico di Roma, via Frascati 33, I-00040 Monteporzio (RM) Italy\\
             \email{carla.maceroni@mporzio.astro.it}
}
			 
   \date{Received ; accepted }
   \abstract{ 
   
   We present and discuss the H$\alpha$ ($\lambda=6563$\AA )  observations of the contact (W UMa type) binaries AE Phoenicis 
and YY Eridani, obtaineded in 1989, 1990 and 1995 with the CAT/CES telescope of the Southern European Observatory (ESO). In 
particular, we compare the intrinsic equivalent widths of both components with the NextGen theoretical models and the 
saturation limit. We find that the average H$\alpha$ equivalent widths are close to the saturation border and that the 
primary components have excess H$\alpha$-emission, indicating enhanced chromospheric activity. This is compatible with both 
theoretical and observational suggestions that the primary is the more magnetically active component and is filled with 
(mostly unresolvable) dark spots and associated chromospheric plages.

   }
	\titlerunning{H$\alpha$ 6563$\AA$ emission in AE Phe and YY Eri}
   \maketitle
%

\section{Introduction}

AE Phoenicis (\object{AE Phe}, G0V, P$_{orb}$ = 0.362 d) and YY Eridani (\object{YY Eri}, G5V, P$_{orb}$ = 0.312 d) are late 
type contact binaries ( W~UMa stars) of W-subtype. Their components touch each other
inside a common convective envelope of constant entropy. According to the theory \citep{lucy}  the primary ( i.e
the more masssive) component should be slightly hotter (by a few hundred degrees) than the companion, but  observations   
show the opposite. 

\citet{mullan} and \citet{rucinski}   ascribed the possible origin of this  discrepancy to
the presence of cool star spots on the primary.  
Some of the evolutionary models predict shallower outer convective zones for the secondary, because of its 
physical status (out of thermal equilibrium), in particular the
angular momentum loss via magnetic braking (AML) models, see e.g. \citet{vilhu82} and \citet{vm88}, and the Thermal 
Relaxation Oscillation models (TRO), see e.g.  \citet{webbink}. 
(Note, by the way,  that \citet{webbink} does not include the AML-models  in his review.)
Shallower convection would mean a less magnetically active secondary,  because for a fixed rotation rate the dynamo action 
is stronger in a thicker convective zone \citep[see, e.g,][]{vilhu87}. 

The observational confirmation of this prediction has remained mostly unexplored although Maceroni et al. (1994),
using both photometry and H$\alpha$ spectroscopy, found a primary slightly cooler than the secondary component and large 
photometric dark spots on the primary surface. Dark spots are generally related to the magnetic (dynamo generated) activity. 

More recently, \citet{barnes}, using high resolution Doppler imaging techniques, found the  primary of AE Phe  
spectroscopically cooler, and provided a further indication of  unresolved dark spots.

In the present paper we re-analyse  the H$\alpha$-observations of AE Phe and YY Eri performed by \citet[][hereafter paper 
I]{maceroni},  together with similar unpublished observations collected in 1995. The motivation is a rejuvenated interest in 
contact binaries, mostly due to the Doppler imaging techniques \citep{barnes}. 
The observations are explained in Section 2 and the H$\alpha$ equivalent widths  compared with  model predictions and with 
the saturation limit (due to the chromospheric emission from plages or from penumbral spot regions filling the photospheric 
absorption). The results are discussed in Section 3 and the conclusions given in Section 4.

 \begin{figure*}
   \centering
\includegraphics[angle=0,width=17cm]{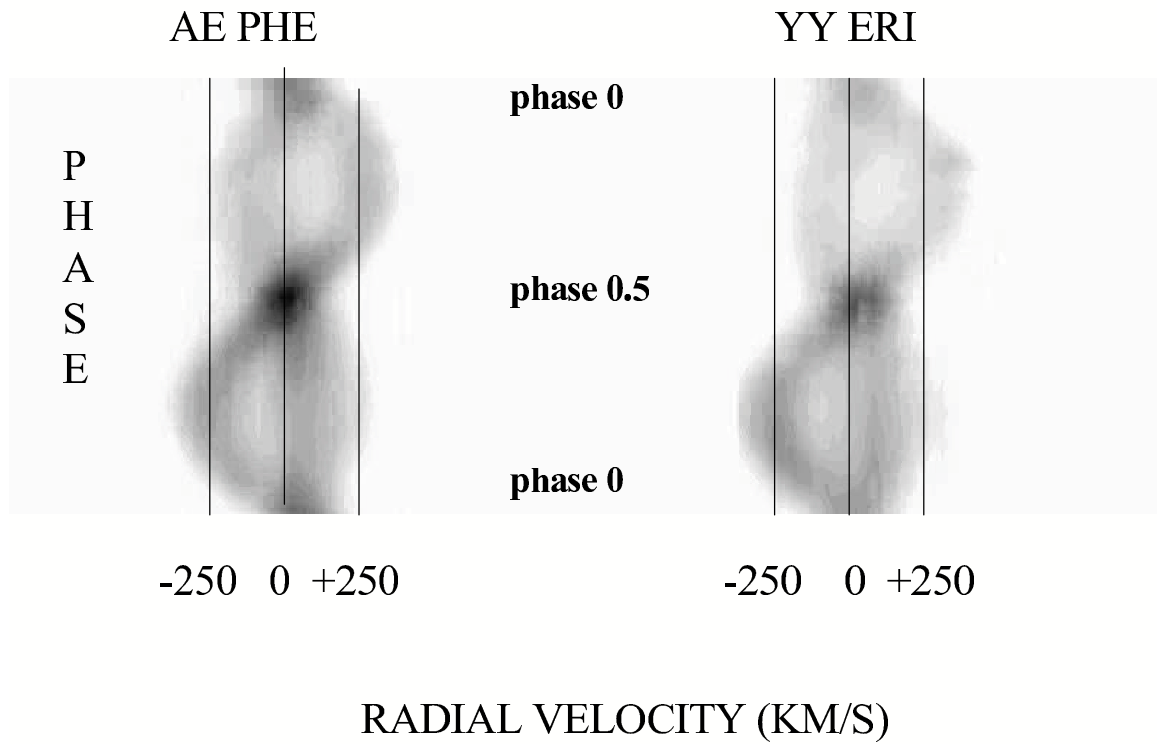}
   \caption{H$\alpha$ 6563$\AA$ dynamic spectra of AE Phe and YY Eri from the 1995 observations. The grey scale is linear, 
the white corresponding to the continuum and the darkest colour (at phase 0.5) to 60 per cent of the continuum level.       
}
              \label{FigGrey}
    \end{figure*}
    \begin{figure*}
   \centering
\includegraphics[angle=0,width=17cm]{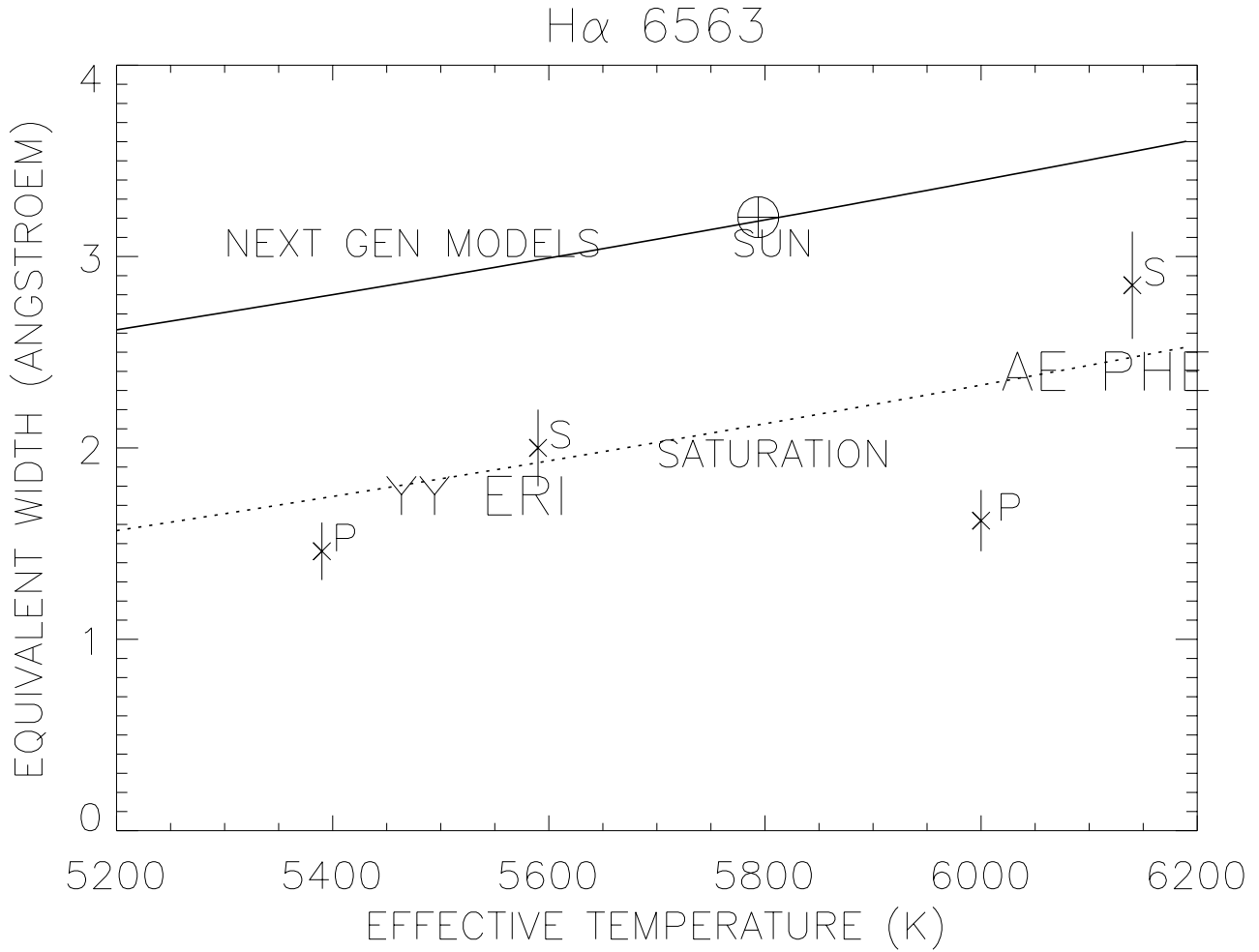}
   \caption{Mean H$\alpha$ equivalent widths  for the primary and secondary components of AE Phe and YY Eri (from Table 1). 
The values from theoretical NextGen-models \citep{hauschildt} are also shown  (solid line) together with  the saturation 
limit found by \citet{herbst} (dotted line). }
              \label{FigEW}
    \end{figure*}


\section{Observations and H$\alpha$ equivalent widths}
 
The observations were performed with the CAT-telescope (Coude Auxiliary Telescope) of the European Southern Observatory 
(ESO at La Silla, Chile) during November 20-25, 1989, November 17-23, 1990, and October 26-31, 1995.  The Coude Echelle 
Spectrometer (CES), with the short camera in the red and resolution of 60 000 ( 5 km s$^-1$) at H$\alpha$ 6563 $\AA$, 
was used. The exposure times were 15 and 20 minutes for AE Phe (m$_V$ = 7.9) and YY Eri (m$_V$ = 8.4), respectively. 
This guaranteed orbital smearing of less than 0.05 in phase for both stars. Complementary photometric observations in B, 
V, and I filters were obtained with the ESO 50 cm telescope. The sky conditions during the 1995 observations, however, 
allowed to get complete, but 
low quality, light curves for AE Phe only. The photometric observations were useful, at any rate,
to check the correct orbital phasing of the spectra.

A sample of line profiles for the years 1989 and 1990 were shown in \citet{maceroni} and are not repeated here for the sake 
of brevity. In Fig. 1  the grey-scale dynamic light curves (phase vs. $\lambda$) for the year 1995 are shown. The grey-scale 
is linear, the white  corresponding to the continuum  and the darkest colour (at phase 0.5) to 60 per cent of the continuum 
level. The radial velocity curves of the broad H$\alpha$-absorption in both components are clearly seen, the curves with 
larger amplitudes corresponding to the secondary (less massive) component.   

 The equivalent widths were measured at elongations, as mean values between phases 0.15 - 0.35 and 0.65 - 0.85.   The 
resulting equivalent widths are, however, relative to the total continuum. Since we are interested in the {\it intrinsic} 
equivalent widths of the components, the measured  values were  further scaled with the  component luminosities. If 
\vspace{0.1cm}
       
       L$_p$ = L/a and L$_s$ = L(a-1)/a, 
       
\vspace{0.1cm}
       
       where L, L$_p$ and L$_s$ are the total, primary and secondary luminosities, respectively, then 
\vspace{0.1cm}
       
      a = 1 + q$^{0.92}$(T$_s$/T$_p$)$^4$ 
\vspace{0.1cm}
       
       which follows directly from the contact condition and the definition of effective temperatures T$_p$ and T$_s$ 
\citep[see e.g.][]{webbink}. Here q is the mass ratio M$_s$/M$_p$. 
       
       Using the photometric effective temperatures and mass ratios found by \citet{maceroni}: ((T$_s$, T$_p$, q) = (6140 K, 
6000 K, 0.39) for AE Phe and = (5590 K, 5390 K, 0.44) for YY Eri), we find a = 1.46 and = 1.54 for AE Phe and YY Eri, 
respectively. These values are very close to those of paper I from photometric solutions.  The intrinsic equivalent widths 
can be computed from the measured ones by 
\vspace{0.1cm}
       
       EW$_p$ = a x EW$_p$(measured) and 
       
       EW$_s$ = a/(a-1) x EW$_s$(measured).

\vspace{0.1cm}    
The equivalent widths are listed in Table 1 and their mean values plotted in Fig.2.    We also used NextGen photospheric 
models\footnote{The NextGen uses the model atmosphere  PHOENIX code.
The  code is available from http://dilbert.physast.uga.edu/yeti},
 with solar abundances and gravity $\log(g) = 4.5$ \citep{hauschildt, short} and computed the H$\alpha$ equivalent widths 
for several temperatures, as shown  in Fig.2.   

These theoretical models match quite well with the solar value marked in Fig.2 (EW = 3.2 $\AA$), as observed with the 
CAT/CES telescope  by exposing  the twilight sky before the observations.  The (absorption) equivalent widths of AE Phe and 
YY Eri are clearly below the theoretical predictions. 

An explanation (which we adopt here) for this deficiency is  that chromospheric emission  fills-in the photospheric 
absorption, thus lowering the measured equivalent widths. \citet{herbst} have estimated this emission for a large sample of 
K-M stars. They found an upper bound (the saturation limit) to the fraction of star's bolometric luminosity that can appear 
as H$\alpha$ emission: L$_{H\alpha}$/L$_{bol}$ = 10$^{-3.9}$. Using F$_{H\alpha}$/F$_{bol}$-values at different effective 
temperatures, as  computed from the  NextGen - models, this relation can be easily converted to the saturation line in 
Fig.2.
 %
\begin{table}
\caption{H$\alpha$ equivalent widths (EW, in units of {\AA}ngstr{\" o}ms) of the more massive (p, primary) and the secondary 
(s) components of AE Phe and YY Eri. The values are average values from the observations of 1989, 1990 and 1995. The 
observed values are direct measurements, with the total luminosity as  continuum, over the orbital phases 0.15 - 0.35 
(marked as 0.25) and 0.65 - 0.85 (0.75).The intrinsic  values are scaled with the components' individual luminosities (see  
text).These intrinsic values are shown in Fig.2. The errors include the differences from epoch to epoch.}
\label{table:1}      
\centering                          
\begin{tabular}{c c c c c}        
\hline\hline                 
comp &EW(0.25)& EW(0.75) & EW  mean & EW  intrinsic \\    %
\hline                        
   AE Phe p & 1.25 $\pm{0.1}$ &1.05 $\pm{0.05}$   &1.15 $\pm{0.07}$  & 1.68 $\pm{0.10}$ \\      
   AE Phe s & 0.90 $\pm{0.07}$ &0.95 $\pm{0.07}$     &0.92 $\pm{0.07}$  & 2.92 $\pm{0.20}$\\
   YY Eri p & 1.00 $\pm{0.10}$ &0.90 $\pm{0.07}$      &0.95 $\pm{0.10}$ &1.45 $\pm{0.15}$\\
   YY Eri s & 0.70 $\pm{0.07}$& 0.75 $\pm{0.07}$    &0.74 $\pm{0.07}$  &2.00 $\pm{0.20}$\\
    
\hline                                   
\end{tabular}
\end{table}
%
 
%
  
%

\section{Discussion}

Both AE Phe and YY Eri clearly  have  equivalent widths of the H$\alpha$-absorption smaller than the Sun and as well than 
those,  predicted by NextGen-models  for normal main sequence stars of similar effective temperatures. This can be 
interpreted 
as being due to extra chromospheric H$\alpha$ emission, that partly fills  the photospheric absorption. The average values 
of both stars lie close to the saturation limit. This behaviour is  similar to other chromospheric emission diagnostics  
\citep[see e.g.][]{vilhu87}, giving additional support for this interpretation.  

The components of YY Eri are not very different from each other, but in AE Phe the primary  has clearly much more 
H$\alpha$-emission than the secondary.  
This is presumably due to a weaker dynamo-generated magnetic activity of the secondary. Since both components rotate with 
the same rate and have almost the same spectral types (effective temperatures) they probably differ with respect to another 
crucial parameter of dynamo theories,  the thickness of the convective zone; the shallower this zone,  the weaker dynamo 
action  \citep[see e.g.][]{vilhu87}.  Theoretical contact binary models  predict  shallower convective zones for the 
secondaries, as well,  due to  their thermal non-equilibrium condition (AML- or TRO-models,  \citep[see e.g.][]{vilhu82, 
webbink}). 

The equivalent widths remained practically the same over all our observing runs, from 1989 to 1990 and 1995. In particular, 
the 1989 and 1990 observations showed that the larger photometric spots are found on the primary star \citep{maceroni}, as 
well as weaker H$\alpha$-absorption, compatible with the present results . \citet{barnes}  interpreted their spectroscopic 
observations (analysed by Doppler-imaging) by introducing unresolved dark spots on the primary. Since the appearance of 
active chromospheres (plages) and cool spots correlate and are the results of the same physical phenomenon, our 
interpretation sounds valid.
   
The phase 0.75 side of the AE Phe primary is  chromospherically more active than the 0.25 side (see Table 1). This is 
compatible with the larger spots found on this side during the first two observing runs  by \citet{maceroni} (paper I).

\section{Conclusions}

We have shown that the contact binaries AE Phe and YY Eri have a weaker photospheric H$\alpha$ 6563 $\AA$ absorption than 
normal slowly rotating main sequence stars of the same spectral type have. This can be interpreted as due to the enhanced 
chromospheric emission in the rapidly rotating components of these contact binaries. This emission is close to the 
saturation limit (see Fig.2) and the  behaviour is similar to many other chromospheric diagnostics  found earlier.  

In  AE Phe the primary (more massive component) is clearly more active in this respect (smaller absorption as compared with 
the secondary or the saturation limit). This is apparently a result from the larger depth of its outer convective zone 
supporting  a stronger dynamo-action, compatible with some structural and evolutionary models of contact binaries:  
AML-models,  \citep[see e.g.][]{vilhu82, vm88} and TRO-models, \citep[see e.g.][for references]{webbink}.
   
\begin{acknowledgements}
      We are grateful to the 6.3 magnitude earthquake during the 1995 observations,  
whose only consequence was that we will remember that night forever.

CM acknowledeges funding of this research project  by MIUR/Cofin and F-INAF programs.
\end{acknowledgements}


\begin{thebibliography}{}

\bibitem[Barnes et al.(2004)]{barnes} Barnes, J.~R., Lister, 
T.~A., Hilditch, R.~W., \& Collier Cameron, A.\ 2004, \mnras, 348, 1321 
\bibitem[Hauschildt et al.(1999)]{hauschildt} Hauschildt, P.~H., 
Allard, F., \& Baron, E.\ 1999, \apj, 512, 377
\bibitem[Herbst \& Miller(1989)]{herbst} Herbst, W., \& 
Miller, J.~R.\ 1989, \aj, 97, 891 
\bibitem[Lucy(1968)]{lucy} Lucy, L.~B.\ 1968, \apj, 151, 
1123 
\bibitem[Maceroni et al.(1994)]{maceroni} Maceroni, C., Vilhu, 
O., van't Veer, F., \& van Hamme, W.\ 1994, \aap, 288, 529 (paper I) 
\bibitem[Mullan(1975)]{mullan}Mullan, D.J., 1975, \apj 198, 563.
\bibitem[Rucinski(1985)]{rucinski}Rucinski,S.M. 1985, in Pringle J.E. and Wade R.A. (eds.), Interacting Binary Stars, 
Cambridge Univ. Press, p. 85.
\bibitem[Short \& Hauschildt(2005)]{short} Short, C.~I., \& 
Hauschildt, P.~H.\ 2005, \apj, 618, 926
\bibitem[Webbink(2003)]{webbink}Webbink R.F. 2003, in Turcotte S., Keller S., Keller S.C., Carallo R.M. (eds.), ASP 
Conf.Ser.Vol.293, 3D Stellar Evolution Contact Binaries, Astron.Soc.Pac., San Francisco, p.76.
\bibitem[van't Veer \& Maceroni(1988)]{vm88} van't Veer, F., 
\& Maceroni, C.\ 1988, \aap, 199, 183 
\bibitem[Vilhu(1982)]{vilhu82} Vilhu, O.\ 1982, \aap, 109, 17 
\bibitem[Vilhu(1987)]{vilhu87}Vilhu, O. 1987, in Linsky J.L. and Stencel R.E. (eds.), Cool Stars, Stellar Systems, and the 
Sun, Lecture Notes in Physics No.291, Springer-Verlag, Berlin-Heidelberg-New York, p.110.
%
%
%
\end{thebibliography}
\end{document}